\documentclass[aps,prl,reprint,superscriptaddress]{revtex4-2}

\usepackage{amsmath}
\usepackage{amssymb}
\usepackage{svg}
\usepackage{siunitx}
\definecolor{bluepoli}{RGB}{0,36,179}
\definecolor{redpoli}{RGB}{204,0,51}
\definecolor{greenpoli}{RGB}{45,137,0}
\definecolor{purplepoli}{RGB}{153,102,204}
\definecolor{azzurropoli}{RGB}{51,53,204}
\definecolor{orangepoli}{RGB}{255,124,17}
\usepackage{hyperref}
\usepackage{physics}
\hypersetup{%
	colorlinks,
	linkcolor={bluepoli}, %
	citecolor={redpoli}, %
	urlcolor={redpoli} %
}

\usepackage{upgreek}

\begin{document}
	
	
	\title{Physics-informed tracking of qubit fluctuations}
	
	\author{Fabrizio~Berritta}
	\affiliation{Center for Quantum Devices, Niels Bohr Institute, University of Copenhagen, 2100 Copenhagen, Denmark}
	
	\author{Jan~A.~Krzywda}
	\affiliation{Lorentz Institute and Leiden Institute of Advanced Computer Science, Leiden University, P.O. Box 9506, 2300 RA Leiden, The Netherlands}
	
	\author{Jacob~Benestad}
	\affiliation{Center for Quantum Spintronics, Department of Physics,
			Norwegian University of Science and Technology, NO-7491 Trondheim, Norway}
	
	\author{Joost~van~der~Heijden}
	\affiliation{QDevil, Quantum Machines, 2750 Ballerup, Denmark}	
	
	\author{Federico~Fedele}
	\affiliation{Center for Quantum Devices, Niels Bohr Institute, University of Copenhagen, 2100 Copenhagen, Denmark}
	\affiliation{Department of Engineering Science, University of Oxford, Oxford OX1 3PJ, United Kingdom}
		
	\author{Saeed~Fallahi}
	\affiliation{Department of Physics and Astronomy, Purdue University, West Lafayette, Indiana 47907, USA}
	\affiliation{Birck Nanotechnology Center, Purdue University, West Lafayette, Indiana 47907, USA}
	
	\author{Geoffrey~C.~Gardner}
	\affiliation{Birck Nanotechnology Center, Purdue University, West Lafayette, Indiana 47907, USA}
	
	\author{Michael~J.~Manfra}
	\affiliation{Department of Physics and Astronomy, Purdue University, West Lafayette, Indiana 47907, USA}
	\affiliation{Birck Nanotechnology Center, Purdue University, West Lafayette, Indiana 47907, USA}
	\affiliation{Elmore Family School of Electrical and Computer Engineering, Purdue University, West Lafayette, Indiana 47907, USA}
	\affiliation{School of Materials Engineering, Purdue University, West Lafayette, Indiana 47907, USA}
	
	\author{Evert~van~Nieuwenburg}
	\affiliation{Lorentz Institute and Leiden Institute of Advanced Computer Science, Leiden University, P.O. Box 9506, 2300 RA Leiden, The Netherlands}
	
	\author{Jeroen~Danon}
	\affiliation{Center for Quantum Spintronics, Department of Physics,
			Norwegian University of Science and Technology, NO-7491 Trondheim, Norway}
	
	\author{Anasua~Chatterjee}
	\affiliation{Center for Quantum Devices, Niels Bohr Institute, University of Copenhagen, 2100 Copenhagen, Denmark}
	
	\author{Ferdinand~Kuemmeth}
	\email{kuemmeth@nbi.dk}
	\affiliation{Center for Quantum Devices, Niels Bohr Institute, University of Copenhagen, 2100 Copenhagen, Denmark}
	\affiliation{QDevil, Quantum Machines, 2750 Ballerup, Denmark}


		
	\date{June 27, 2024}
	
	\begin{abstract}
Environmental fluctuations degrade the performance of solid-state qubits but can in principle be mitigated by real-time Hamiltonian estimation down to time scales set by the estimation efficiency. 
We implement a physics-informed and an adaptive Bayesian estimation strategy and apply them in real time to a semiconductor spin qubit. 
The physics-informed strategy propagates a probability distribution inside the quantum controller according to the Fokker--Planck equation, appropriate for describing the effects of nuclear spin diffusion in gallium-arsenide. Evaluating and narrowing the anticipated distribution by a predetermined qubit probe sequence enables improved dynamical tracking of the uncontrolled magnetic field gradient within the singlet-triplet qubit. 
The adaptive strategy replaces the probe sequence by a small number of qubit probe cycles, with each probe time conditioned on the previous measurement outcomes, thereby further increasing the estimation efficiency.  
The combined real-time estimation strategy efficiently tracks low-frequency nuclear spin fluctuations in solid-state qubits, and can be applied to other qubit platforms by tailoring the appropriate update equation to capture their distinct noise sources.
	\end{abstract}
	
	\maketitle
	
	\section{Introduction}
	
	Low-frequency environmental fluctuations cause decoherence in solid-state qubits ~\cite{Chirolli2008, Bergli2009, Falci2024}. Quantum error correction strategies \cite{Terhal2015} can detect and correct errors but demand an increased number of physical qubits. Conventional noise reduction techniques, such as dynamical decoupling \cite{Viola1999, Szankowski2017} and active suppression of environmental fluctuations~\cite{Foletti2009, London2013, Scheuer2016}, are not universally effective and may not align with specific experimental goals.
	
Hamiltonian learning emerges as a promising solution for compensating for uncontrolled environmental effects and enhancing the qubit quality factor \cite{Shulman2014,Nakajima2020,Kim2022, Vepsaelaeinen2022, Yun2023, Berritta2024, Park2024}. This approach leverages modern hardware capabilities to provide real-time feedback, but comes at the cost of dedicating time to estimate the fluctuating Hamiltonian parameters. Although several theoretical estimation schemes \cite{Sergeevich2011, Cappellaro2012,Ferrie2013, Bonato2017, Scerri2020, GutierrezRubio2020, Craigie2021, Benestad2023} have been proposed to boost the estimation efficiency, no experiment has yet demonstrated a \emph{physics-informed} scheme within any qubit platform, where understanding of the physical processes driving the fluctuations is utilized to improve the estimations. 	
Even the experimental adoption of real-time \emph{adaptive} Bayesian strategies~\cite{Bonato2015, Arshad2024}, where measurement parameters are chosen based on the previous measurements, is still missing in gate-defined spin qubits. 
This work reports the first real-time physics-informed and adaptive Bayesian estimation of a qubit.
	
To demonstrate an adaptive and physics-informed estimation protocol, we employ a singlet-triplet (ST$_0$) qubit in GaAs. 
In nitrogen-vacancy centers in diamond \cite{Schirhagl2014} and semiconductor spin qubits \cite{Burkard2023}, low-frequency noise from spinful nuclear isotopes decreases qubit performance through hyperfine interactions. Isotopic purification techniques \cite{Feher1958, Gordon1958} mitigate this issue in group IV semiconductors such as silicon and germanium, though it comes with significant effort and does not remove low-frequency noise originating from other sources. 
For our demonstration we chose GaAs as its nuclear noise spectrum is well understood~\cite{Taylor2007,Malinowski2017}. Our technique involves programming a commercial quantum controller, powered by an integrated field-programmable gate array (FPGA), to propagate the probability distribution of the effective nuclear fields on the dots in real time, using the Fokker--Planck (FP) equation~\cite{Craigie2021, Benestad2023}. This enables the dynamic tracking of the fluctuating nuclear field gradient across the qubit, which is the main source of decoherence in ST$_0$ qubits in GaAs~\cite{Taylor2007, Malinowski2017}.
	
The propagation of probability distributions on the quantum controller, here according to the FP equation, can be replaced by other update equations, e.g., a transition matrix for Markov processes~\cite{Asmussen2003}, or machine-learning-based methods for signal prediction~\cite{Gupta2018}, the details of which depend on the specific nature of the qubit system. 

Real-time capabilities of quantum controllers can also be used advantageously to choose optimal measurement parameters within an \emph{adaptive} estimation sequence (in our case updating free-induction-decay times on the fly, based on previous measurement outcomes), which we will analyze separately below. 

Our scheme can accomplish Hamiltonian learning for intermittent calibration of circuit parameters, making it ideal for the recurrent and reliable execution of quantum circuits against the impact of drift. Since the interleaved estimation and qubit operation take place in the same qubit, there is an intricate interplay between the correlation time of the fluctuations being estimated, the efficiency of the estimation procedure, and the required time for coherent operations between estimations. Optimizing and managing these time scales will be essential when going from single-qubit devices to multi-qubit devices, making it even more valuable to estimate qubit and noise correlations times quickly and efficiently.
	\section{Results}
	
	\subsection{Device and  Bayesian estimation}
	\begin{figure}
		\includegraphics{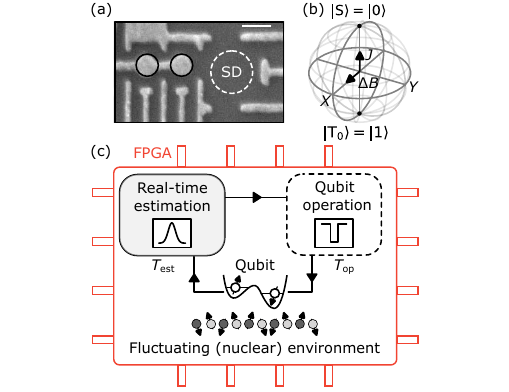} 
\caption{\label{fig:Fig1}\textbf{Qubit implementation and estimation schedule}. 
\textbf{(a)} Scanning-electron micrograph of a GaAs double-dot device similar to the one used in this work~\cite{Fedele2021} comprising a singlet-triplet qubit (black circles) next to a sensor dot (SD) used for qubit readout. Scale bar $\SI{100}{\nano\meter}$.
			\textbf{(b)} 
			Exchange coupling $J(\varepsilon)$ and Overhauser gradient $\Delta B\propto hf_B(t)$ drive rotations of the qubit around two orthogonal axes of the Bloch sphere, providing universal qubit control if the prevailing Overhauser frequency $f_B$ can be estimated sufficiently efficiently. 
			\textbf{(c)} 
Qubit schedule, alternating between periods $T_{\text{op}}$ of quantum information processing (dashed box), and short periods $T_{\text{est}}$ for efficiently learning the fluctuating environment (gray box).
		}
	\end{figure}
	
We employ the top-gated GaAs double quantum dot (DQD) array from~\cite{Fedele2021} with one of its ST$_0$ qubits activated using gate electrodes as in Figure~\ref{fig:Fig1}(a). 
A dilution refrigerator provides a base temperature below $\SI{50}{\milli\kelvin}$ and a 200-mT in-plane magnetic field defines the $z$-direction. 
A commercial DAC~\cite{QDAC} (FPGA-powered quantum controller~\cite{QM}) applies low-frequency (high-frequency) baseband waveforms to the gate electrodes, and radio-frequency reflectometry off one ohmic contact of the sensor dot distinguishes the charge configurations of the DQD, allowing single-shot qubit readout~\cite{Vigneau2023}. 
Details about the experimental setup can be found in~\cite{Berritta2024}. 
	
The qubit operates in the $(1,1)$ and $(0,2)$ charge configuration, where the integers stand for the number of electrons in the left and right dot of the DQD.
	In the two-electron ST$_0$ basis, the Hamiltonian can be approximated in the regime of interest as
	\begin{equation}
		\mathcal{H}(t) = \frac{J(\varepsilon)}{2}\sigma_z + \frac{g^*\mu_{\text{B}} \Delta B(t) }{2}\sigma_x,
	\end{equation}
	where $\sigma_i$ represent the Pauli operators, $g^*$ is the effective $g$-factor, and $\mu_{\text{B}}$ is the Bohr magneton.
	The energy $J(\varepsilon)$ characterizes the exchange interaction between the two electrons, which is tunable via the relative electrical detuning of the dots. By defining $\varepsilon=0$ at the (1,1)--(0,2) charge-state degeneracy, detuning is proportional to the difference in the effective on-site potentials on the two dots of the singlet-triplet qubit, where negative $\varepsilon$ corresponds to the (1,1) ground-state region. 
	The field $\Delta B(t)$ denotes the $z$-component of the Overhauser gradient, which is the difference in effective magnetic fields on the two dots due to the hyperfine interaction of the electrons with approximately $10^5$--$10^6$ of spinful nuclei on each dot~\cite{Taylor2007}.
	This gradient fluctuates slowly, and our goal is to efficiently estimate the corresponding Overhauser frequency $f_B(t) \equiv g^* \mu_{\rm B} \Delta B(t)/h$ in real time on the quantum controller, using a physics-informed model with and without adaptive probe times. 
	
A Bloch-sphere representation of the two contributions to ${\cal H}$ is sketched in Fig.~\ref{fig:Fig1}(b).
The qubit undergoes manipulation through voltage pulses applied to the plunger gates of the DQD, which effectively control the magnitude of $J(\varepsilon)$. Deep in the (1,1) regime, where $J(\varepsilon) \ll |hf_{B}|$, the qubit is almost purely driven by the Overhauser gradient, whereas close to $\varepsilon=0$ typically $J(\varepsilon) \gtrsim |hf_{B}|$.
	
	After manipulation, the qubit is measured by projecting the unknown final spin state onto either the (1,1) charge state ($|\text{T}_0\rangle$) or the (0,2) charge state ($|\text{S}\rangle$), by tuning to positive $\varepsilon$. Each single-shot readout of the DQD charge configuration involves the generation, demodulation, and thresholding of a few-microsecond-long radio-frequency burst on the quantum controller~\cite{Berritta2024}.
	
	The fluctuating frequency $f_B$ is assessed on the quantum controller using a Bayesian estimation approach based on a series of $N$ free-induction-decay experiments with evolution times $t_i$ where $i=1,2,\dots, N$~\cite{Shulman2014,Nakajima2020, Kim2022, Yun2023, Berritta2024, Park2024}. Employing $m_i$ to represent the outcome ($|\text{S}\rangle$ or $|\text{T}_0\rangle$) of the $i$-th measurement, the likelihood function $P(m_i | f_B)$ is defined as the probability of obtaining $m_i$ given a value of $f_B$,
	\begin{equation} \label{eq:bayes_1}
		P\left(m_i | f_B\right)=\frac{1}{2}\left[1+m_i\left(\alpha+\beta \cos \left(2 \pi f_B t_i\right)\right)\right],
	\end{equation}
	where $m_i$ takes a value of 1 ($-1$) if $m_i = \ket{\text{S}}$ ($\ket{\text{T}_0}$), and $\alpha$ and $\beta$ are parameters accounting for the measurement error and axis of rotation on the Bloch sphere during a free-induction decay experiment~\cite{Shulman2014}. In this work we use $\alpha = 0.28$ and $\beta =0.45$ extracted from a series of separate free-induction decay (FID) experiments.
	Applying Bayes' rule to estimate $f_B$ based on the series of measurements $m_N, \ldots m_1$, which are assumed to be independent of each other,  yields the final probability distribution $P_{\text{final}}\left(f_B\right) \equiv P\left( f_B\ | m_N, \ldots m_1\right)$ given by
	\begin{equation}
		\begin{aligned}
			P_{\text{final}}\left(f_B\right) \propto P_0\left( f_B\right)  \prod_{i=1}^N\left[1+m_i\left(\alpha+\beta \cos \left(2 \pi f_B t_i\right)\right)\right],
		\end{aligned}
	\end{equation}
	where $P_0\left(f_B\right)$ is the initial probability distribution assumed for $f_B$ before the estimation starts. Equivalently, the measurement outcome $m_i$ updates the Bayesian probability distribution according to $P_{i}(f_B) \propto P_{i-1}(f_B) P(m_i | f_B)$, up to a normalization factor, where the likelihood function $P(m_i | f_B)$ is given by Eq.~(\ref{eq:bayes_1}).
	The final estimate of $f_B$ is taken to be the expectation value $\langle f_B \rangle$, calculated over the final distribution $P_{\text{final}}\left(f_B\right)$ after all $N$ measurements have been performed. The estimation protocol can be repeated at user-defined times when the qubit is not in use for other operations. 
	
	Estimating low-frequency fluctuations is useful as outlined in the following example, depicted in Figure~\ref{fig:Fig1}(c):
	one starts by estimating the instantaneous magnitude of the slowly fluctuating field (the Overhauser frequency in our case), resulting in a strongly reduced uncertainty in this field.
	Subsequently, that knowledge is used to compensate for the random value of the field during coherent qubit operation, resulting in an increased qubit quality factor~\cite{Berritta2024}.
	However, while operating the qubit for a period $T_{\text{op}}$, the field will again slowly drift, which can be captured by letting its distribution function evolve over time according to a known noise model~\cite{Malinowski2017}. 
For the Overhauser gradient, this amounts to a diffusion of its mean towards zero mean field and an increase of the uncertainty towards a maximum value that depends on the number and coupling strengths of the involved nuclear spins. 
Before such a stationary state is reached, the known dynamics of the probability distribution can be used to improve the feedback or make the next estimation more efficient.
After a user-defined period $T_{\text{op}}$, qubit operations are momentarily halted and a new real-time estimation is initiated on the quantum controller. Its duration, approximately $T_{\text{est}}\propto N$, depends on the desired estimation accuracy as discussed below. A series of estimation sequences, each resulting in an accurate distribution $P_{\text{final}}(f_B)$, is what we refer to as qubit tracking.  
	
	\subsection[Non-tracking and physics-informed tracking of the qubit frequency]{Physics-informed tracking of the qubit frequency}
	
This section describes how such ``stroboscopic'' physics-informed tracking of an Overhauser field is implemented on the quantum controller and to what extent it produces higher-quality estimates than obtainable via more commonly used estimation sequences~\cite{Shulman2014,Berritta2024}.
The protocol is physics-informed in the sense that the assumed evolution of the distribution function in between two estimations is based on a physical model describing the nuclear spin dynamics in GaAs-based quantum dots.

	\begin{figure*}
		\includegraphics{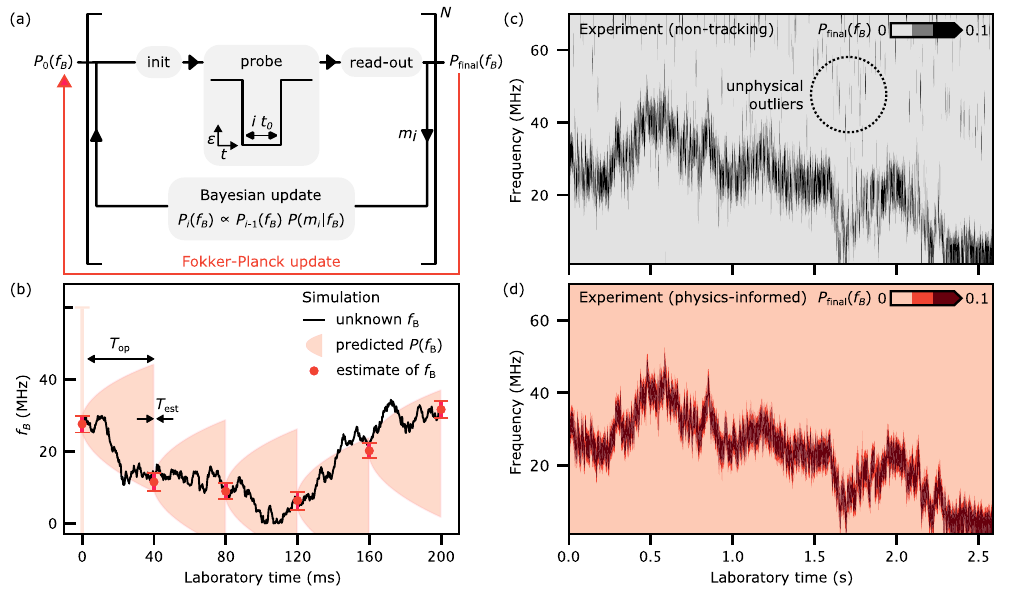} 
		\caption{\label{fig:Fig2} 
\textbf{Tracking of the Overhauser frequency by anticipating nuclear spin diffusion on the quantum controller.}
\textbf{(a)} 
The physics-informed estimation sequence for $f_B$ initializes the prior distribution $P_0(f_B)$ by evolving an older final distribution $P_{\text{final}}(f_B)$ (Fokker-Planck update). For each of the $N$ probe cycles, labeled $i$, the quantum controller initializes the qubit to the singlet state, performs a FID for time $t_i = i t_0$, then updates the probability distribution $P_{i}(f_B)$ based on the measurement outcome $m_i$. After $N$ probe cycles, the final distribution $P_{\text{final}}(f_B)$ is saved.
\textbf{(b)} 
Simulation of the unknown fluctuating Overhauser gradient (black) and five physics-informed estimation sequences, illustrating the tracking protocol. Every $\SI{40}{\milli\second}$, a sequence of FID probe cycles results in a final distribution with expected value $f_B^f = \langle f_B\rangle$ and error bar $2\sigma_f$ (red markers). 
The simulation assumes a uniform prior distribution $P_0(f_B)$ at $t=0$, whereas subsequent priors $P_0(f_B)$ are based on the mean $\mu(t)$ and standard deviation $\sigma(t)$ propagated by the Fokker-Planck equation over period $T_\text{op}$ (shaded in light red). 
\textbf{(c)}
Experimental results for the non-tracking reference protocol, using  $P_0(f_B) \equiv P_{\text{uniform}}(f_B)$ for each estimation sequence.
\textbf{(d)}
Experimental results for the physics-informed tracking protocol, obtained simultaneously with non-tracking estimates in panel (c). 
The initial prior $P_0(f_B)$ for each column is $P_{\text{final}}(f_B)$ from the previous column, propagated in time according to Equation~\eqref{eq:gaussian_parameters}. Note the absence of multi-peaked distributions $P_{\text{final}}(f_B)$. 
		}
	\end{figure*}
	
	The FPGA-based estimation of the Overhauser frequency $f_B$ is illustrated in Fig.~\ref{fig:Fig2}(a):
	One estimation sequence consists of $N$ repetitions of a free-induction decay (FID) probe cycle.
	In each probe cycle, a singlet pair is initialized in (0,2) and then detuned deeply into the (1,1) region. At $\varepsilon \approx \SI{-40}{\milli\volt}$, the quantum controller lets the qubit evolve for a probe time $t_i = i \, t_0$, before thresholding the resulting qubit state and updating the probability distribution [$P_{i}(f_B) \propto P_{i-1}(f_B) P(m_i | f_B)$]. 
In this sequence, the probe times $t_i$ are predetermined and linearly distributed by the probe time spacing $t_0 = \SI{1}{\nano\second}$. We assume that $N$ is sufficiently small such that the Overhauser gradient remains constant during the sequence.

We model the dynamics of the Overhauser gradient as an Ornstein--Uhlenbeck (or drift--diffusion) process~\cite{Malinowski2017}, driven by randomly occurring nuclear spin flips. 
The time dependence of the distribution function $P(f_B,t)$ resulting from such a process is governed by a Fokker--Planck (FP) equation~\cite{kampenStochasticProcessesPhysics1990,Craigie2021,Benestad2023}, allowing the prediction of $P(f_B)$ in periods when the qubit is used for other operations ($T_\text{op}$). 
Assuming that each final distribution $P_{\text{final}}(f_B)$ is sufficiently characterized by its mean and variance, we instruct the  quantum controller to approximate it by a Gaussian distribution~\footnote{Since we only work with positive frequencies $f_B$, one should be aware that this becomes inaccurate for distributions that have significant weight close to $f_B=0$, i.e., that have a variance larger than the square of the mean. }.
Denoting the mean and variance immediately after estimation (time $t=0$) as $f_B^f$ and $\sigma_f^2$, respectively, the FP equation yields as solution for $t>0$
	\begin{equation}\label{eq:gaussian}
		P(f_B, t)=\frac{1}{\sqrt{2 \pi \sigma(t)^2}} \exp \left\{-\frac{[f_B-\mu(t)]^2}{2 \sigma(t)^2}\right\},
	\end{equation}
	where
	\begin{subequations}\label{eq:gaussian_parameters}
		\begin{eqnarray}
			\mu(t) &=& f_B^f e^{-\Gamma t}, \label{eq:gaussian_parameters_average}  
			\\
			\sigma(t)^2&=&\sigma_K^2+\left[\sigma_f^2-\sigma_K^2\right] e^{-2 \Gamma t}. \label{eq:gaussian_parameters_sigma}
		\end{eqnarray}
	\end{subequations}
Here, $\sigma_K$ is the steady-state root mean square value of the Overhauser field frequency (typically around $30-\SI{50}{\mega\hertz}$~\cite{Malinowski2017}), while $\Gamma$ reflects the slow relaxation rate of nuclear spin polarization (measured to be $\Gamma \approx \SI{1.1}{\hertz}$ from autocorrelation). Notably, the inverse of $\Gamma$, denoted as $T_{\text{c}} = \Gamma^{-1} \approx \SI{0.91}{\second}$, defines the timescale for the correlation of fluctuations in $f_B$; this establishes the time window within which an estimate of $f_B$ is expected to remain useful. 

In Figure~\ref{fig:Fig2}(b) we numerically simulate a fluctuating Overhauser gradient with $T_{\text{c}} = \SI{1}{\second}$ and $\sigma_K = \SI{30}{\mega\hertz}$. The associated unknown frequency $f_B$ (black trace) is assumed to be estimated every $T_{\text{op}}=\SI{40}{\milli\second}$ (red markers). 
The physics-informed evolution of probability distributions (shaded red areas, adapted from Ref.~\cite{Benestad2023}) captures two properties expected for nuclear spin diffusion, namely the inclination of the average of the Overhauser gradient to drift back towards zero [Eq.~\eqref{eq:gaussian_parameters_average}], and a progressive expansion of the uncertainty in the gradient towards $\sigma_K$ [Eq.~\eqref{eq:gaussian_parameters_sigma}]. 
Both processes take place on a timescale of $T_{\text{c}}$. 

Initially, no knowledge of $f_B$ is available, reflected by a uniform prior distribution $P_{\text{uniform}}(f_B)$ at $t=$0 represented by the semi-transparent error bar spanning the entire frequency range of the simulation ($\SI{60}{\mega\hertz}$)~\footnote{We choose our prior distribution to be non-zero for positive frequencies only, resulting in a unimodal final distribution. As the sign of the Overhauser gradient is unknown, the true final distribution would always be symmetric around zero.}. 
A number of FID cycles are performed until the updated probability distribution has a fitted $\sigma < \SI{2}{\mega\hertz}$. 
This estimation sequence is assumed to take only a few hundred microseconds, i.e. much shorter than $T_{\rm op}$ and $T_{\text{c}}$, and we only plot the mean $f_B^f$ and 95\% confidence interval of the final distribution $P_{\text{final}}(f_B)$ (first red marker). 

After the first estimation sequence, the Overhauser fields are left to evolve freely for $T_{\rm op}$.
During this time, the distribution function is assumed to be Gaussian; the time dependence of its mean and variance is given by Eq.~\eqref{eq:gaussian_parameters}. 
The evolution of the 95\% confidence interval is indicated by the red shaded area in Fig.~\ref{fig:Fig2}(b). At the end of $T_{\rm op}$ ($t=\SI{40}{\milli\second}$) the associated Gaussian distribution is characterized by $\mu(T_{\rm op})$ and $\sigma(T_{\rm op})$, and serves as the initial prior distribution for the next estimation sequence. 
Similarly, estimations at $t=80$, 120, 160 and 200~ms use as prior the most recent Gaussian. 
	
If $T_{\text{op}}$ is smaller than $T_{\text{c}}$, the physics-informed $P_0(f_B)$ will remain somewhat constrained, providing a better prior compared to a uniform distribution and potentially requiring fewer FID experiments for a more accurate estimate of $f_B$. 
If $T_{\text{op}}$ becomes comparable to or larger than $T_{\text{c}}$, prior knowledge about $f_B$ becomes irrelevant and is not expected to improve the next estimation. 
		
To experimentally test the benefits of physic-informed priors, we define a non-tracking estimation scheme that always sets the initial distribution $P_0(f_B)$ to a uniform distribution $P_{\text{uniform}}(f_B)$ between $\SI{1}{\mega\hertz}$ and $\SI{70}{\mega\hertz}$ with $\SI{1}{\mega\hertz}$ resolution.
Thus, as in previous works~\cite{Shulman2014,Berritta2024}, each estimation sequence does not retain any memory of previous estimations. 
In parallel to this non-tracking estimation, we instruct the quantum controller to also generate estimates based on the physics-informed initialization of $P_0(f_B)$, thereby improving the estimation accuracy as quantified below.  
	
Figure~\ref{fig:Fig2}(c) plots 1,000 final probability distributions of the non-tracking scheme, acquired over a span of \SI{2.6}{\second} using an $N=31$ schedule with $T_{\text{est}}=\SI{0.6}{\milli\second}$ and $T_{\text{op}}=\SI{2}{\milli\second}$. 
Specifically, each FID probe cycle lasts $\SI{20}{\micro\second}$, of which $\SI{5}{\micro\second}$ is dedicated to qubit readout, $\SI{2.6}{\micro\second}$ to initialize the qubit and discharge the bias tee with a zero-averaging pulse, and the remaining time is used to update the non-tracking and physics-informed distributions $P_{i}(f_B)$ on the FPGA. 
Several estimation sequences result in a multi-peaked probability distribution, with secondary peaks that randomly jump from one column to another. 
In simulations, such ``outliers" also appear in the absence of measurement errors and appear to be a shortcoming of the algorithm, not an artifact of the device or the quantum controller.  
The known correlation time of the Overhauser field dynamics makes it improbable that the sudden jumps of the outliers represent the actual Overhauser field gradient, and similar jumps in previous work were associated with compromised qubit quality factors (cf. discussion of Fig.~2b of Ref.~\cite{Berritta2024} in its supplementary information).  	
	
Figure~\ref{fig:Fig2}(d) shows the physics-informed estimates $P_{\text{final}}(f_B)$, acquired concurrently with the non-tracking estimates in panel (c). 
Strikingly, multi-peaked probability distributions are absent, suggesting that the physics-informed model on the quantum controller suppresses unphysical jumps of the estimated Overhauser gradient (here with $T_{\text{c}} = \SI{0.91}{\second}$ and $\sigma_K = \SI{50}{\mega\hertz}$). 
By extracting the standard deviation from each column in Figure~\ref{fig:Fig2}(d), we find that its average is reduced relative to the average standard deviation extracted from panel~\ref{fig:Fig2}(c), suggesting an improved estimation accuracy. 

	\begin{figure}
		\includegraphics{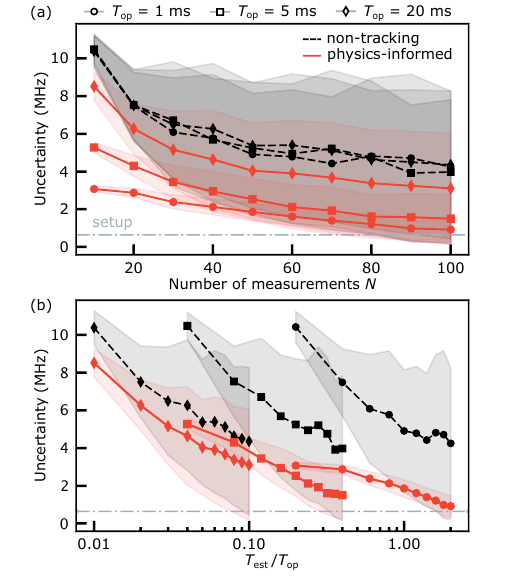} 
		\caption{\label{fig:Fig3}\textbf{Efficiency of the non-tracking and physics-informed protocols.}
\textbf{(a)} Estimation uncertainty as a function of the number of FID probes in the estimation sequence, for the non-tracking (black) and physics-informed (red) protocols. 
Symbols denote the average standard deviation of 10,000 $\langle f_B\rangle$ values, whereas shaded regions show their standard deviation, for different choices of operation time. 
\textbf{(b)} Uncertainty from (a) plotted as a function of the ratio $T_{\text{est}}/T_{\text{op}}$, where the estimation time is $T_{\text{est}} = N\cdot\SI{20}{\micro\second}$. 
The dash-dotted gray line indicates the resolution limit imposed by our setup, see main text.  
		}
	\end{figure}
	
Figure~\ref{fig:Fig3} compares the performance of the non-tracking and physics-informed estimation sequences as a function of the number of FID probe cycles, for different choices of $T_{\text{op}}$. Each data point corresponds to an independent experiment comprising 10,000 repetitions of an estimation sequence. 
The plotted uncertainty is defined as the average standard deviation of the final probability distribution $P_{\text{final}}(f_B)$ of each of the 10,000 estimations. The shaded areas indicate the standard deviation of the associated 10,000 standard deviations. 
In our experiment, the true value of the real field, and thus the actual error in the estimation, is unknown, and therefore we rely on the uncertainty measure plotted as a reasonable metric. Indeed, low uncertainties at the end of $T_\text{est}$ correlate with increased quality factors of controlled Overhauser rotations during $T_\text{op}$ (see Supplemental Material~\cite{supplementary}). 
	
The uncertainty of the non-tracking estimates in Fig.~\ref{fig:Fig3}(a) does not depend on $T_{\text{op}}$. This is expected, as the prior distributions $P_0(f_B)$ in the non-tracking scheme are always the uniform distribution $P_{\text{uniform}}(f_B)$, with no memory of the previous estimates. 
In contrast, the uncertainty of the physics-informed estimates decreases with decreasing $T_{\text{op}}$, for fixed number of measurements in the estimation sequence. This suggests that a narrower prior yields a more accurate estimate. 
	
Remarkably, with as few as 10 probes the physic-informed estimates for $T_{\text{op}}=\SI{1}{\milli\second}$ are more accurate than non-tracking estimates based on 100 probes (in Fig.~\ref{fig:Fig3}(a) the uncertainties are approximately 3~MHz and 5~MHz, respectively.)
With increasing number of probe cycles, the uncertainty of non-tracking estimates saturates near $\SI{5}{\mega\hertz}$, whereas the physics-informed estimation uncertainty approaches the limitation imposed by our choice of frequency binning ($\SI{0.8}{\mega\hertz}$~\footnote{We programmed the frequency resolution on the quantum controller to be $\SI{1}{\mega\hertz}$. The associated minimum standard deviation of $P_{\text{final}}(f_B)$ calculated on the FPGA is approximately 0.8~MHz.}).

The trade-off between ``qubit duty cycle" ($T_\text{op}/T_\text{est}$) and estimation accuracy is evident in Fig.~\ref{fig:Fig3}(b). Here, we replot the uncertainties from (a) as a function of the estimation time $T_{\text{est}} = N\cdot \SI{20}{\micro\second}$, where $N$ is the number of qubit probes and $\SI{20}{\micro\second}$ is the probe cycle duration. Depending on the desired Hamiltonian uncertainty, a maximum operation limit $T_\text{op}$ and a significant qubit downtime (high $T_\text{est}/T_\text{op}$ ratio) for estimation must be tolerated. 
The optimum choice of $N$ depends on details of the noise spectrum and the estimation efficiency~\cite{Benestad2023}.

One may be tempted to pursue the lowest possible uncertainty while estimating the environmental fluctuations, but the operational benefits will depend on details such as the tolerable estimation uncertainty for a certain application and how long it is expected to survive given a specific environment. Because achieving lower uncertainties in general requires more qubit down time for estimation, quantum information processing applications may need to define a tolerated ``error budget'', which translates into a useful operation time $T_{\rm op}$ depending on the correlation time of the fluctuations $T_{\text{c}}$ and a minimized estimation time $T_{\rm op}$ depending on the efficiency of the protocol. 
	
	So far, we demonstrated an improved Hamiltonian learning protocol that tracks a slowly fluctuating environmental parameter, by instructing a quantum controller to generate in real time physics-informed priors. Next, we instruct the controller to adaptively choose the probe times, thereby reducing the length of the estimation sequences. 	
	\subsection[Physics-informed adaptive Bayesian tracking of the qubit frequency]{Adaptive Bayesian tracking of the qubit frequency}
	
	\begin{figure*}
		\includegraphics{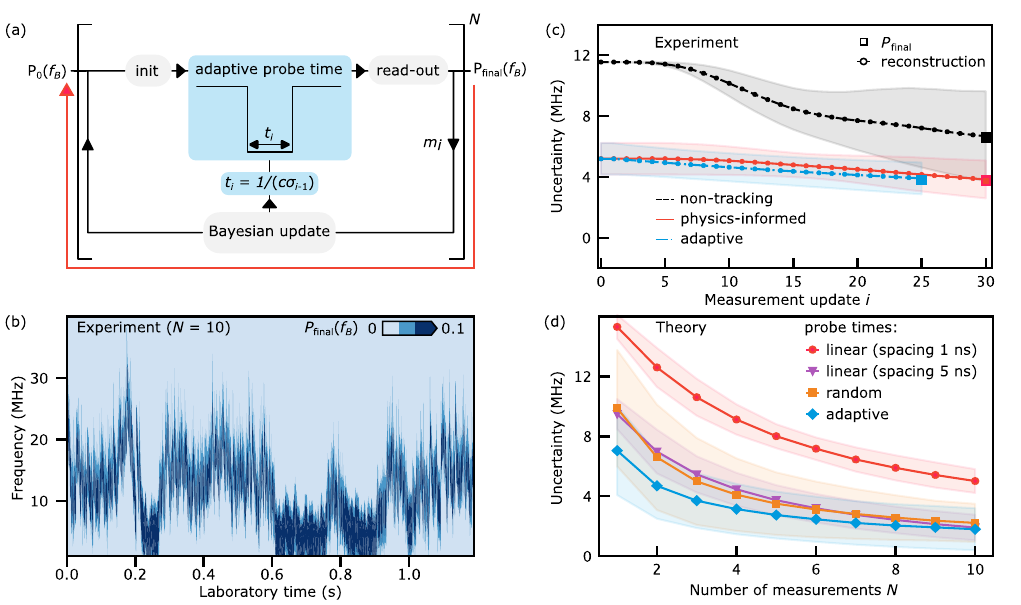} 
		\caption{\label{fig:Fig4}
\textbf{Adaptive Bayesian tracking by real-time choice of qubit probe times.}
\textbf{(a)} 
In this adaptive Bayesian estimation sequence, probe times $t_i$ are chosen based on the standard deviation $\sigma_{i-1}$ of the previous Bayesian distribution. $P_0(f_B)$ is initialized based on the FP equation. 
\textbf{(b)} 
Adaptive tracking obtained from short estimation sequences ($N=10$) for $T_{\rm op} = \SI{1}{\milli\second}$. 
\textbf{(c)} Reconstructed uncertainty in the distribution function within an estimation sequence (defined in the text) as a function of the measurement update $m_i$. Squares at the end of the curves correspond to the experimental posterior distributions computed on the quantum controller.
\textbf{(d)} Simulated uncertainty expected at the end of a short estimation sequence ($N$$\leq$10) for different probe time protocols, including evenly distributed $t_i$ (probe time spacing of 1 or 5 ns), adaptive probe times, and random probe times (see main text). The initial prior distributions are assumed to be determined from the FP equation.
		}
	\end{figure*}
	
For the purpose of only monitoring fluctuating Hamiltonian parameters without interspersed qubit operation, non-adaptive Bayesian estimation is straightforward to execute because it does not require real-time feedback and could even be carried out a posteriori. 
However, numerical studies \cite{Sergeevich2011, Cappellaro2012, Bonato2017, Scerri2020, Craigie2021, Benestad2023} suggest the beneficial use of adaptive estimation sequences in which the probe times $t_i$ are chosen based on previous measurement outcomes,
as experimentally realized in nitrogen-vacancy centers~\cite{Bonato2015,Arshad2024}. 
	
Previous experiments with gate-defined spin qubits employed non-tracking and non-adaptive FID-based Bayesian estimation to probe the qubit frequency~\cite{Shulman2014,Berritta2024}. 
In this section, we supplement the generation of physics-informed time-evolved priors by the generation of adaptive probe times in real time, thereby reducing the number of required probes and showing a path towards much shorter estimation sequences.  

Figure~\ref{fig:Fig4}(a) illustrates the key difference of the adaptive estimation sequence, relative to that in Fig.~\ref{fig:Fig2}(a): the free-evolution time $t_i$ for the $i$-th FID probe now depends on the previous Bayesian update as
	\begin{equation}
		t_i=\frac{1}{c\sigma_{i-1}}, \label{eq:csigma}
	\end{equation}
where $\sigma_{i-1}$ is the standard deviation of the Gaussian-approximated probability distribution $P_{i-1}(f_B)$, except $\sigma_{0}$, which is the standard deviation of prior $P_0(f_B)$ based on the FP equation. 
The optimal numerical prefactor $c$ is expected to depend on the experimental setup~\cite{Scerri2020}. 
Intuitively, this choice for the free evolution times can be motivated by our desire that two oscillations with frequencies that differ by $\Delta f$ develop a phase shift of $\pi$ after time $t = 1/(2\Delta f)$. In other words, Eq.~(\ref{eq:csigma}) maps a frequency range of width $c \sigma_{i-1}/2$ to a large phase contrast in the likelihood function. 
	
Implementation of the estimation protocol of Fig.~\ref{fig:Fig4}(a) on the quantum controller yields reliable estimates for $f_B$ from only $10$ probes per sequence, as shown in Fig.~\ref{fig:Fig4}(b) for $T_{\rm op} = \SI{1}{\milli\second}$ and $c\approx 13$~\footnote{Due to the numerical precision of the quantum controller and the discreteness of the $t_i$ that can be implemented, the actual ratio between $1/\sigma_{i-1}$ and $t_i$ varies slightly between FID probe cycles. }. 
This example demonstrates the estimation of a slowly fluctuating qubit frequency within 200 microseconds, which is one order of magnitude shorter and with better accuracy than previously reported~\cite{Berritta2024}. 
Here, $c\approx 13$ was chosen empirically, and further improvements may be possible by better choices informed from numerical simulations, see the Supplemental Material~\cite{supplementary}.

Outliers appear to be absent both for the physics-informed [Fig.~\ref{fig:Fig2}(c)] and adaptive tracking [Fig.~\ref{fig:Fig4}(b)], likely for similar reasons, motivating a quantitative comparison based on experimental data and theoretical insights. 
	
Figure~\ref{fig:Fig4}(c) compares average uncertainties, inferred from experimental data in Fig.~S2 of the Supplemental Material~\cite{supplementary}. 
We choose $T_\text{op}=\SI{5}{\milli\second}$ and perform 10,000 repetitions of three protocols, focusing on $N$$\leq$30 to test whether short sequences benefit from adaptive probe cycles. 
The three squares at the end of the curves show the uncertainties $\sigma$ (defined as in Fig.~\ref{fig:Fig3} and computed on the quantum controller from the posterior distributions $P_{\text{final}}$) for non-tracking (black), physics-informed (red), and adaptive (blue) estimation sequences. For $N=30$, the non-tracking scheme yields an average $\sigma\approx\SI{7.3}{\mega\hertz}$, while the physics-informed scheme yields $\sigma\approx\SI{3.5}{\mega\hertz}$. The uncertainty of the adaptive scheme is similar, though obtained with fewer probes ($N=25$).
	
To investigate how each probe cycle contributes information gain, we analyze how uncertainties evolve within a sequence (additional details can be found in Fig.~S2~\cite{supplementary}). Specifically, we reconstruct the Bayesian probability updates $P_{i}(f_B)$ from our record of raw single-shot measurement outcomes $m_i$
~\footnote{To increase the estimation bandwidth within the FPGA memory constraints, the quantum controller overwrites Bayesian updates $P_{i}(f_B)$ and only records $P_{\text{final}}(f_B)$ at the end of each sequence, as well as all $N$ measurement outcomes.}. 
For each $i$, we plot the standard deviation of $P_{i}(f_B)$ (reconstruction), averaged over all 10,000 repetitions, as well as their standard deviation (shaded areas). 

The non-tracking method is clearly outperformed by the physics-informed and adaptive schemes. This is expected, as both the physics-informed and adaptive protocols use physics-informed prior distributions. Furthermore, the adaptive scheme has consistently lower uncertainty than the physics-informed scheme, though only marginally. Finally, we note that the uncertainties for the non-tracking and physics-informed schemes barely decrease during the first few measurements ($i\lesssim 5$), as shown by the nearly flat curves in this range. In contrast, the adaptive scheme shows a negative slope already for the first measurement outcomes, indicating information gain and a narrowing of the probability distribution. 
		
To explore the ultimate estimation efficiencies that can be expected for our spin-qubit system, unconstrained by coarse frequency binning and limited memory on the FPGA-powered controller, we now turn towards simulated Overhauser fluctuations, assumed to follow an Ornstein--Uhlenbeck process with $T_{\text{c}} = \SI{1}{\second}$ and $\sigma_K = \SI{40}{\mega\hertz}$, and simulate estimation sequences on a much finer and larger frequency grid (0 to 150~MHz with 0.25~MHz bin size) than currently possible in our experimental setup.  
	
Figure~\ref{fig:Fig4}(d) shows the resulting uncertainties and their standard deviations, assuming $T_\text{op}=\SI{5}{\milli\second}$, for different distributions of probe times (see Figs.~S3,~S4~\cite{supplementary} for further details):
In the sequences with ``linear" probe times, $t_i=i\, t_0$, we observe that the choice of the probe time spacing $t_0$ (shown 1 and 5~ns) has a drastic influence on the resulting accuracy. 
In the sequences with ``random" probe times, $t_i$ is randomly chosen from a uniform distribution between $\SI{1}{\nano\second}$ and $\SI{50}{\nano\second}$. 
In the sequences with ``adaptive" probe times, $t_i=1/(c\sigma_{i-1})$, now with $c=6$ and without rounding $t_i$ to the temporal granularity of the quantum controller (see the Supplemental Material~\cite{supplementary}). 

The adaptive-probe-time sequence outperforms the linear sampling approach with $t_0 = \SI{1}{\nano\second}$, yielding uncertainties that are on average smaller by a factor of $\approx2.7$, and is also superior to $t_0 = \SI{5}{\nano\second}$ and random probe times, resulting in approximately 30\% smaller uncertainties for short estimation sequences ($N\lesssim 5$). 
We therefore believe that adaptive estimation sequences will become crucial in applications that only permit a small number of probe cycles. 
	
In summary, the results shown in Fig.~\ref{fig:Fig4} present the first adaptive Bayesian estimation scheme implemented in a semiconductor-defined spin qubit.
	
	The real-time capabilities of the quantum controller enable probe times $t_i$ to be updated based on previous measurement outcomes $m_{i-1},m_{i-2},\dots, m_1$, resulting in a small but measurable improvement compared to linearly spaced probe times. 
	Our approach is substantiated by numerical simulations, indicating that high-quality estimates of the qubit frequency achieving only a few percent error (approximately $\SI{3}{\mega\hertz}$ uncertainty with a simulated dynamic range of $\approx\SI{150}{\mega\hertz}$) should be possible with fewer than five qubit probe cycles.  
	
	\section{Outlook}
	
We have implemented physics-informed and adaptive estimation sequences that allowed the efficient tracking of low-frequency fluctuations in a solid-state qubit. A quantum controller estimates in real time the uncontrolled magnetic field fluctuations in a gallium-arsenide singlet-triplet spin qubit, yielding improved accuracy by temporally evolving a sufficiently recent probability distribution according to the Fokker-Planck equation. In addition, the adaptive choice of qubit probe times, based on the standard deviation of the updated probability distribution, allows for significantly shorter estimation sequences yielding similar or reduced uncertainties. Compared to previous experiments~\cite{Berritta2024}, this work extends the estimation bandwidth from a few hundred Hz to $\approx\SI{2.5}{\kilo\hertz}$, due a tenfold reduction of the estimation time and a reduced uncertainty. 
	
While our work marks the first real-time adaptive tracking of a semiconductor spin qubit, determining optimal protocols compatible with constraints of the control hardware and application requirements remains an open question. We anticipate further progress by research that combines theoretical and hardware aspects.  
	
Possibly useful modifications of the protocol could relax the assumption of single-shot readout~\cite{Dinani2019} or mitigate state preparation and measurement errors by duplication of probe cycles~\cite{Sergeevich2011, Ferrie2013}. Probe times can further be optimized by also taking into account the estimated qubit frequency, not just its uncertainty, and possibly it is advantageous to terminate an estimation sequence when reaching an accuracy target, rather than a predetermined length. 

Fault-tolerant quantum computing based on quantum error correction will likely require qubits that are affected by limited amounts of Markovian noise. Therefore, real-time frequency tracking protocols may become important tools, as they suppress non-Markovian noise~\cite{Park2024}.
 
	By properly modifying the tracking equation relevant to the specific noise source, this work offers an efficient, physics-informed, and adaptive Hamiltonian learning protocol for real-time estimation of low-frequency noise in solid-state qubits.

	\section{Author contributions}
	F.B. led the measurements and data analysis, and wrote the manuscript with input from all authors. F.B., J.v.d.H., A.C. and F.K. performed the experiment with theoretical contributions from J.A.K., J.B., J.D. and E.v.N. F.F. fabricated the device.	S.F., G.C.G. and M.J.M. supplied the heterostructures. A.C. and F.K. supervised the project.
	
	\section{Acknowledgments}
	We thank Torbj\o{}rn Rasmussen for help with the experimental setup.
	This work received funding from the European Union's Horizon 2020 research and innovation programme under grant agreements 101017733 (QuantERA II) and 951852 (QLSI), from the Novo Nordisk Foundation under Challenge Programme NNF20OC0060019 (SolidQ), from the Inge Lehmann Programme of the Independent Research Fund Denmark, from the Research Council of Norway (RCN) under INTFELLES-Project No 333990, as well as from the Dutch National Growth Fund (NGF) as part of the Quantum Delta NL programme. 

	\nocite{*}
	\bibliography{my_bibliography}
\end{document}


\beginsupplement
	
	
	
	\title{Supplemental Material for ``Physics-informed tracking of qubit fluctuations''}
	
	\author{Fabrizio~Berritta}
	\affiliation{Center for Quantum Devices, Niels Bohr Institute, University of Copenhagen, 2100 Copenhagen, Denmark}
	
	\author{Jan~A.~Krzywda}
	\affiliation{Lorentz Institute and Leiden Institute of Advanced Computer Science, Leiden University, P.O. Box 9506, 2300 RA Leiden, The Netherlands}
	
	\author{Jacob~Benestad}
	\affiliation{Center for Quantum Spintronics, Department of Physics,
		Norwegian University of Science and Technology, NO-7491 Trondheim, Norway}
	
	\author{Joost~van~der~Heijden}
	\affiliation{QDevil, Quantum Machines, 2750 Ballerup, Denmark}	
	
	\author{Federico~Fedele}
	\affiliation{Center for Quantum Devices, Niels Bohr Institute, University of Copenhagen, 2100 Copenhagen, Denmark}
	\affiliation{Department of Engineering Science, University of Oxford, Oxford OX1 3PJ, United Kingdom}
	
	\author{Saeed~Fallahi}
	\affiliation{Department of Physics and Astronomy, Purdue University, West Lafayette, Indiana 47907, USA}
	\affiliation{Birck Nanotechnology Center, Purdue University, West Lafayette, Indiana 47907, USA}
	
	\author{Geoffrey~C.~Gardner}
	\affiliation{Birck Nanotechnology Center, Purdue University, West Lafayette, Indiana 47907, USA}
	
	\author{Michael~J.~Manfra}
	\affiliation{Department of Physics and Astronomy, Purdue University, West Lafayette, Indiana 47907, USA}
	\affiliation{Birck Nanotechnology Center, Purdue University, West Lafayette, Indiana 47907, USA}
	\affiliation{Elmore Family School of Electrical and Computer Engineering, Purdue University, West Lafayette, Indiana 47907, USA}
	\affiliation{School of Materials Engineering, Purdue University, West Lafayette, Indiana 47907, USA}
	
	\author{Evert~van~Nieuwenburg}
	\affiliation{Lorentz Institute and Leiden Institute of Advanced Computer Science, Leiden University, P.O. Box 9506, 2300 RA Leiden, The Netherlands}
	
	\author{Jeroen~Danon}
	\affiliation{Center for Quantum Spintronics, Department of Physics,
		Norwegian University of Science and Technology, NO-7491 Trondheim, Norway}
	
	\author{Anasua~Chatterjee}
	\affiliation{Center for Quantum Devices, Niels Bohr Institute, University of Copenhagen, 2100 Copenhagen, Denmark}
	
	\author{Ferdinand~Kuemmeth}
	\email{kuemmeth@nbi.dk}
	\affiliation{Center for Quantum Devices, Niels Bohr Institute, University of Copenhagen, 2100 Copenhagen, Denmark}
	\affiliation{QDevil, Quantum Machines, 2750 Ballerup, Denmark}	
	
	
	\date{June 27, 2024}
	\maketitle
	\tableofcontents
	\section{Variance of the distribution as a measure of estimation error}
	
	\begin{figure}[h]
		\centering
		\includegraphics[width=\textwidth]{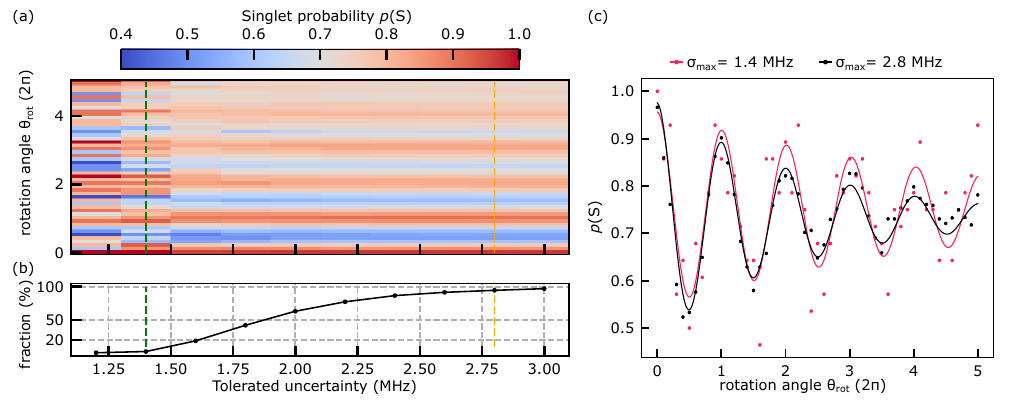}
		\caption{\textbf{Improved qubit quality factor from lower estimation uncertainties.}
			\textbf{(a)} Average measured singlet--triplet oscillations resulting from averaging over only the best estimates, as a function of rotation angle and cutoff uncertainty $\sigma_\text{est,max}$.
			\textbf{(b)} The fraction of used estimations as a function of $\sigma_\text{est,max}$.
			\textbf{(c)} Dots: the averaged oscillations at $\sigma_\text{est,max} = \SI{1.4}{\mega\hertz}$ (red, corresponding to rejecting 97\% of the repetitions), and $\SI{2.8}{\mega\hertz}$ (black, rejecting 3\%), corresponding to the green and orange vertical dashed lines in (a,b), respectively.
			Solid lines: Fitted exponentially decaying sinusoidal oscillations.
		}
		\label{fig:FigS1}
	\end{figure}
	The true frequency of the Overhauser field gradient is not known, so to benchmark the different estimation protocols we choose as a figure of merit the standard deviation $\sigma_{\text{est}}$ of the final probability distributions resulting from the estimations.
	Formally, this standard deviation follows from
	\begin{equation}
		\sigma_{\text{est}}^2 = \text{Var}(f_B) \equiv \sum_n \big(f_B[n] - \langle f_B \rangle\big)^2 P_{\text{final}}(f_B[n]),
	\end{equation}
	where the index $n$ labels the bins of the (discrete) probability distribution stored on the quantum controller.
	In this section we demonstrate that lower $\sigma_{\text{est}}$ indeed correlates with better estimation of $f_B(t)$, by performing Overhauser-driven controlled rotations of the qubit based on the information provided by the final probability distributions~\cite{Berritta2024}.
	
	We thus perform a series of $N=3,668$ estimations of the Overhauser gradient, spanning a few seconds of laboratory time. The experiment ends whenever the number of repetitions with Overhauser-controlled rotations (explained later) reaches $1,000$. Due to finite FPGA program memory, the controlled rotations are executed whenever $\SI{20}{\mega\hertz}\le \langle f_B \rangle \le \SI{45}{\mega\hertz}$. The repetitions of controlled rotations happen to be executed $1,000$ times out of the $N=3,668$ repetitions of estimations because of the chosen frequency range.
	For each estimation repetition,	the quantum controller initializes the qubit in the singlet state, pulses deep into the (1,1) region, where the qubit undergoes rotations along $\sigma_x$ with the instantaneous frequency $f_B(t)$, and after time $\tau_{\rm rot}$ the quantum controller pulses back into the (0,2) region for qubit readout.
	The quantum controller repeats this experiment $3,668$ times with $\tau_{\rm rot}$ linearly increasing by 1 ns from $\tau_{\rm rot} = \SI{1}{\nano\second}$ to $\tau_{\rm rot} = \SI{50}{\nano\second}$. From the 50 single-shot outcomes of each estimation repetition, $\langle f_B \rangle $ is estimated.
		
	After each estimation repetition, if $\SI{20}{\mega\hertz}\le \langle f_B \rangle \le \SI{45}{\mega\hertz}$, the quantum controller performs Overhauser-driven controlled rotations of the qubit by an user-defined unit less target angle $\theta_{\rm rot} = 2\pi \langle f_B \rangle \tilde{\tau}_{\rm rot}$, where $\langle f_B \rangle$ is the expectation value for $f_B$ resulting from the estimation performed just before the cycle of rotation experiments, which is different for each trace, $\theta_{\rm rot}$ is linearly spaced between 0 and 5 in 51 points and $\tilde{\tau}_{\rm rot}$ is computed on-the-fly on the quantum controller.
	The controlled rotations consist of the following steps:
	the quantum controller initializes the qubit in the singlet state, pulses deep into the (1,1) region, where the qubit undergoes rotations along $\sigma_x$ with the instantaneous frequency $f_B(t)$, and after time $\tilde{\tau}_{\rm rot}$ to rotate the qubit by the wanted angle $\theta_{\rm rot}$, the quantum controller pulses back into the (0,2) region for qubit readout.
	The time $\tilde{\tau}_{\rm rot}$ thus corresponds exactly to the required time to rotate the qubit by the user-defined angle of rotation $\theta_{\rm rot}$, if the frequency was indeed exactly $\langle f_B \rangle$.
	We then study the average of the 1,000 traces based on their respective estimation uncertainty $\sigma$; the quality factor of the averaged oscillations is the result of the qubit decoherence time and the average accuracy of the knowledge about $f_B$. We assume the qubit decoherence time does not change across repetitions and the quality factor is mostly dependent on the uncertainty on $f_B$.
	
	To show that low-quality estimates play an important role in the loss of oscillation amplitude we post-process the measured data based on the $\sigma_\text{est}$ of all repetitions:
	We introduce the variable $\sigma_\text{est,\text{max}}$, and for given $\sigma_\text{est,\text{max}}$ we reject all repetitions with a probability distribution with a calculated variance of $\sigma^2_\text{est} > \sigma^2_\text{est,\text{max}}$, and we average over the remaining traces.
	We consider the controlled rotations of the qubit taken whenever the $\SI{20}{\mega\hertz}\le \langle f_B \rangle \le \SI{45}{\mega\hertz}$ (the chosen interval is limited by the quantum controller program memory).
	In Fig.~\ref{fig:FigS1}(a,b) we show the result for 10 choices of $\sigma_\text{est,\text{max}}$, ranging from $\SI{1.2}{\mega\hertz}$ to $\SI{3.0}{\mega\hertz}$.
	Fig.~\ref{fig:FigS1}(a) shows the singlet probability $P(S)$ of the averaged oscillations as a function of $\tau$ and $\sigma_\text{est,\text{max}}$, and in Fig.~\ref{fig:FigS1}(b) we plot the corresponding fraction of used data for each $\sigma_\text{est,\text{max}}$.
	In Fig.~\ref{fig:FigS1}(c) we focus on two specific choices for $\sigma_\text{est,\text{max}}$ [$\SI{1.4}{\mega\hertz}$ (red) and $\SI{2.8}{\mega\hertz}$ (black)], that correspond to rejecting $97\%$ and $3\%$ of the estimations, respectively [see the vertical dashed lines in Fig.~\ref{fig:FigS1}(a,b)].
	The dots represent the averaged oscillations, as shown in Fig.~\ref{fig:FigS1}(a), and the solid curves fitted sinusoidal oscillations with an exponentially decaying envelope.
	We thus see a significant improvement in the quality factor of the oscillations, which suggests that an important part of the observed decay may be associated with the performance of the estimation scheme.
	In that sense, $\sigma_{\text{est}}$ thus seems to be a valid metric to benchmark the different protocols when the real field is not known.

	\clearpage
	\section{Reconstructing the distribution during estimation [Fig.~4(\lowercase{c}) of main text]}
	
	\begin{figure}[h]
		\centering
		\includegraphics[width=\textwidth]{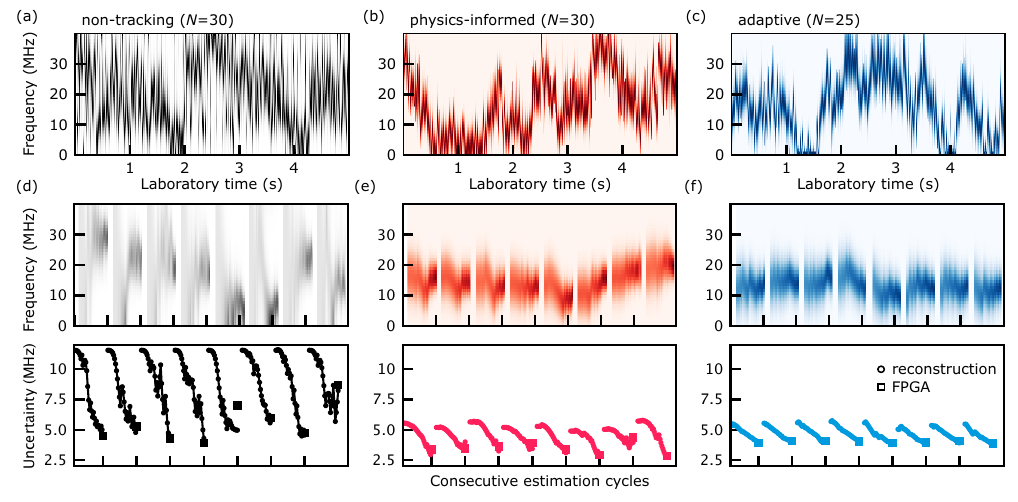}
		\caption{\textbf{Reconstruction of probability distributions.}
			\textbf{(a--c)}  Final probability distribution on the quantum controller of 1,000 repetitions, using the non-tracking (a), tracking (b), and adaptive scheme (c).
			\textbf{(d--f)} Reconstructed evolution of the distribution during the estimation, for 8 consecutive repetitions (top panels) and corresponding reconstructed uncertainty $\sigma$ during estimation (bottom panels).
		}
		\label{fig:FigS2}
	\end{figure}
	
	In Fig.~\ref{fig:FigS2} we present more detailed data underlying Fig.~4(c) in the main text.
	Figure~\ref{fig:FigS2}(a--c) presents the first 1,000 out of 10,000 repetitions of experimental final posterior distributions $P_{\text{final}}(f_B)$ computed on the quantum controller with $\sigma_K=\SI{40}{\mega\hertz}$ for the estimation methods tracking (a), tracking (b) (both with a total number of measurements $N=30$), and adaptive (c) (with $N=25$). These repetitions serve as the benchmark for comparing the different methods in Fig.~4(c) of the main text and the following. To compare the methods fairly in Fig.~4(c) of the main text we exclude estimation runs where the estimated frequency $\langle f_B\rangle$ is smaller than the minimum measurable value of $\approx\SI{2.5}{\mega\hertz}$ resulting from discretization in the quantum controller program (such small $\langle f_B\rangle$ cause problems in the adaptive-time scheme, as explained in more detail in Fig.~\ref{fig:FigS4x} below). 
	In the top row of Fig.~\ref{fig:FigS2}(d--f) we display the reconstructed evolution of the probability distributions during the estimation procedure for 8 consecutive repetitions, where the ticks at the horizontal axis mark the end of each estimation. The bottom row shows the corresponding evolution of the uncertainty $\sigma$ during the estimations.
	While the distributions shown in (a--c) have been computed on the quantum controller, (d--f) show results that were reconstructed from the recorded measurement outcomes $m_i$ (see description below). As in the main text, squares represent uncertainties computed from the distribution in (a), demonstrating good agreement with the reconstructions most of the time, with a few exceptions. These occasional deviations are likely attributed to variations in numerical accuracy between the quantum controller and the desktop computer. A detailed discussion of setup limitations is provided in the next section.
	
	\subsection{Method of reconstructing from the measurement outcomes}
	
	The reconstruction process involves the analysis of raw data, obtained from experimental measurements by the quantum controller in conjunction with the evolution times $t_i$ used, where $i = 1,2\ldots N$ labels measurement updates (shots). As in the main text, we label the repetitions of estimation cycles by \(n\). It follows the evolution times $t_{n,i}$ are loaded from the quantum controller. On top of this, we use the corresponding thresholded reflectometry measurement. As a result, the matrix representing single-shot measurements $m_{n,i} = \pm 1$ is generated. 
	
	Subsequently, the elements of the arrays \(t_{n,i}\) and $m_{n,i}$ are employed to update the probability distribution based on a Bayesian update rule:
	\[
	P_{n,i}(f_{B,j}) = \mathcal{N}_{n,i}P_{n,i-1}(f_{B,j}) [1+m_{n,i}(\alpha+\beta \cos(2 \pi f_{B,j} t_{n,i}))]/2,
	\]
	where \(\mathcal N_{n,i}^{-1} = \sum_j P_{n,i-1}(f_{B,j}) [1+m_{n,i}(\alpha+\beta \cos(2 \pi f_{B,j} t_{n,i}))]/2\) and \(f_{B,j}\) is defined on a discrete grid.
	
	At each pair of indices \(n,i\), the first two moments are computed:
	
	\[
	\langle (f_B)^k \rangle_{n,i}  = \sum_{j} P_{n,i}(f_{B,j}) (f_{B,j})^k,
	\]
	which can be linked to the expectation value (\(\mu_{n,i}\)) and the standard deviation (\(\sigma_{n,i}\)) of the field:
	\[
	\mu_{n,i} = \langle f_B \rangle_{n,i}, \quad \sigma_{n,i} =\sqrt{ \langle f_B^2 \rangle_{n,i}  - \langle f_B \rangle_{n,i}^2}.
	\]
	
	The quantity \(\sigma_{n,i}\) serves as a measure of field uncertainty, which is used as a figure of merit in the study. For visual representation, reconstructed values of \(\sigma_{n,i}\) are depicted in Fig.~\ref{fig:FigS2} d-f) as a function of \(i\) for eight consecutive realizations \(n = 641,642
	\ldots 648\).
	
	In the first repetition (\(n=1\)), the initial distribution \(P_{1,0}(f_B)\) is flat. In subsequent repetitions (\(n>1\)), either a flat distribution (non-adaptive schemes) or a Gaussian distribution with parameters computed from the previous $n-1$  are used (adaptive schemes). For the adaptive prior methods, we use the update equation:
	\[
	P_{n+1,0}'(f_{B,j}) = \exp\left(-\frac{(f_{B,j}-\mu_{n+1,0}[T_{op}])^2}{2 \sigma_{n+1,0}^{2}   [T_{\text{op}}]}\right), 
	\]
	with the normalization: 
	\[
	\quad P_{n+1,0}(f_{B,j}) = P_{n+1,0}'(f_{B,j})/\sum_{j'} P_{n+1,0}'(f_{B,j'}),
	\]
	where the expectation value and uncertainty are propagated from the last measurement of the
	previous estimation sequence, using Fokker-Planck update:
	
	\[
	\mu_{n+1,0}[T_\text{op}] =   \mu_{n,N} e^{-T_\text{op}/T_c}, \quad \sigma_{n+1,0}^{2}[T_{op}] = \sigma_K^2 + \left(\sigma_{n,N}^2 - \sigma_K^2\right)e^{-2T_\text{op}/T_c}.
	\]
	\(T_\text{op}\) serves as the separation time between two consecutive repetitions \(n\) and \(n+1\). In the reconstruction we use parameters that mimic experimental setup, i.e. we set $\alpha = 0.28$, $\beta = 0.45$, $\sigma_K = \SI{40}{\mega\hertz}$, $T_c \approx \SI{1} {\second}$ and use the frequency grid \({f_{B,j} = 1,2\ldots \SI{40}{\mega \hertz}}\)
	
	\clearpage
	\section{Numerical simulation of estimations}
	
	We support our results by the numerical study, that aims at simulating relevant features of estimation setup. For each realization of the algorithm we generate a random trajectory of the field \(f_B(t)\), modeled as the Ornstein-Uhlenbeck process with \(\sigma_K = \SI{40}{\hertz}\) and \(T_c \approx \SI{1}{\second}\). We use quasistatic approximation, i.e. assume that the frequency is constant during single evolution time $f_{B}^{(r,i)}$ and varies only between the consecutive FID experiments. For each repetition $r$ and each measurement $i$, we use the value of simulated field to compute probability of measuring singlet $P_{r,i}(f_B^{(r,i)}) = 1+(\alpha+\beta \cos(2 \pi f_B^{(r,i)} t)]/2$, where the evolution time $t$ is selected based on the method used. With this distribution, we use a random number generator to draw the single-shot outcomes $m_{r,i}$, which we then feed to the algorithm described in the previous section.
	
	\subsection{Determine optimal adaptive-time parameter $c$}
\begin{figure}[h]
		\centering
		\includegraphics{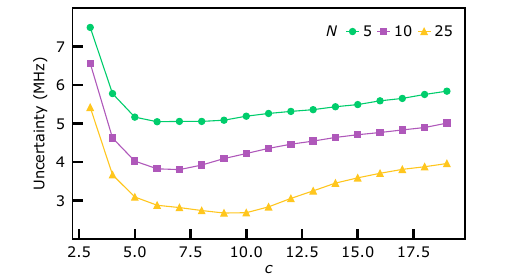}
		\caption{\textbf{Optimal choice of $c$ in $t_i= 1/(c\sigma_{i-1})$.} Numerically simulated uncertainty as a function of $c$ and number of measurements $N  = 5$ (green), $N= 10$ (blue), $N=25$ (yellow) for the adaptive estimation method. To reflect experimental scenario we set $\alpha = 0.28$, $\beta = 0.45$, $\sigma_K = \SI{40}{\mega\hertz}$, $T_c \approx \SI{1} {\second}$ and us the frequency grid $f_j = 1,2\ldots \SI{40}{\mega \hertz}$. The estimation protocol consists of 1,000 realizations, and is averaged over 100 independent numerical experiments. }
		\label{fig:FigS3x}
	\end{figure}

	Firstly, we use the simulation to find the optimal value of parameter $c = 1 / (t_i\sigma_{i-1})$, that minimizes estimation uncertainty, given the constraints of the quantum controller. To achieve that, we kept the simulation parameters from experimental and reconstruction protocols, the details of which are described in the previous section. We sweep value of $c$ in the numerical simulation of the adaptive method that uses $N=5,\,10,\,25$ measurements (colors) and find that $c \approx 5-10$ typically gives the smallest uncertainties [see Fig.~\ref{fig:FigS3x}].
	
As our simulations do not include other sources of qubit dephasing, this range of $c$ may not be the best choice in the experiment. In principle, as $\sigma_{i-1}$ decreases, longer sensing times are more efficient in further decreasing the frequency uncertainty. In practice, however, the range of useful sensing times is limited by the qubit decoherence time, on the order of $\SI{100}{\nano\second}$ in our system. Consequently, for experiments with $N=25$, we empirically tuned the parameter $c$ to $c\approx 13$, which seems to limit the allowed sensing time to a reasonable range.  
	
	\subsection{Benchmarking the different estimation methods [Fig.~4(d) of main text]}
	
	In our experimental setup, the estimations encountered limitations stemming from readout quality and quantum controller memory constraints. We envision that advancements in both aspects could significantly enhance the speed and robustness of the adaptive-time scheme (see also next section).
	To explore the potential, we conduct a benchmark of the estimation schemes outlined in the main text, comparing them against two additional schemes incorporating adaptive priors and using distinct evolution time schedules: randomly picked times $t_i$ and linearly spaced $t_i$ but with a larger time step.
	We thus simulate a series of 4,000 estimations spaced by $T_{\rm op} = \SI{5}{\milli\second}$, spanning $\SI{20}{\second}$ of ``lab'' time, using different estimation schemes. In our estimations, the true Overhauser gradient follows an Ornstein--Uhlenbeck process with $\sigma_K = \SI{40}{\mega\hertz}$ and $T_c \approx \SI{1}{\second}$.
	We further assume for simplicity ideal conditions, amounting to:
	\begin{itemize}
		\item ideal readout ($\alpha = 0$, $\beta = 1$);
		\item a broader frequency grid $[0,150]\,\SI{}{\mega\hertz}$ instead of $(0,40]\, \SI{}{\mega\hertz}$ with $\SI{0.25}{\mega\hertz}$ bin size;
		\item the possibility to use non-integer sensing times ($t_i= i t_0$ with $t_0=1$~ns).
	\end{itemize}
	We compute the average uncertainty at the end of 4,000 estimation cycles, separated by $T_\text{op} = \SI{5}{\milli\second}$. For statistical purpose we additionally average this quantity over 100 independent realizations of the field and measurement outcomes. Obtained in this way average uncertainty, as a function of number of measurements in each estimation $N$, is shown in Fig.~4d) in the main text. 
	
	\begin{figure}[h]
		\centering
		\includegraphics[width=\textwidth]{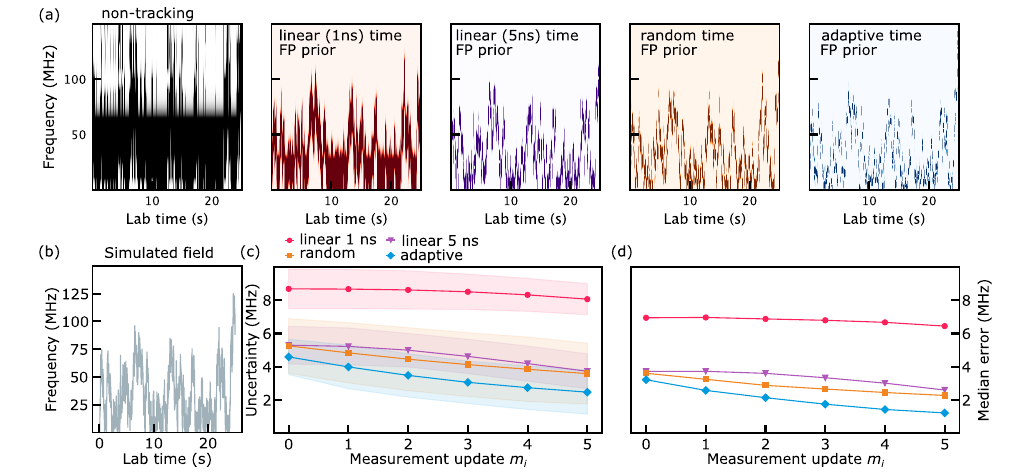}
		\caption{\textbf{Numerical simulation of estimation methods.}
			\textbf{(a)} Simulated posterior distributions $P_{\text{final}}(f_B)$ for different estimation methods, given the simulated frequency shown in (b). ``FP prior" means that the prior distribution $P_{\text{0}}(f_B)$ at the beginning of each repetition follows from the Fokker--Planck equation, as described in the main text.
			\textbf{(b)} Simulated frequency of the Overhauser magnetic field gradient.
			\textbf{(c)} Statistics of the uncertainties in the final distributions of the protocols shown in (a) (except the inefficient non-tracking protocol), as a function of the measurement updates $i = 0,1,2,3,4,5$ in estimation cycle [same plot as Fig.~4(c) in the main text].
			\textbf{(d)} Median of the absolute error of the protocols shown in (a) (except non-tracking) as a function of $i = 0,1,2,3,4,5$ [same legend as in panel (c)].
		}
		\label{fig:FigS4x}
	\end{figure}
	
	To shed more light on the performance of different estimation schemes as well as the correlation between uncertainty and the error we concentrate on a single realization of the field.  Fig.~\ref{fig:FigS4x}(a) shows the resulting final distributions as a function of the lab time for $N=5$, where all plots are based on the same simulated realization of $f_{B,{\rm sim}}(t)$, whose trajectory is shown in Fig.~\ref{fig:FigS4x}(b).
	The statistics of the uncertainties in the final distributions $P_{\rm final}(f_B)$ (their mean and standard deviation) are plotted in Fig.~\ref{fig:FigS4x}(c) as a function of measurement update, where value $i=0$ corresponds to average initial distribution while $i=5$ is the average final one. 
	We again used the uncertainties in $P_{\rm final}(f_B)$ as a measure for the error in the estimate, in order to make comparison to the experimental results fair.
	However, in the simulations we of course know the true instantaneous value $f_{B,{\rm sim}}(t)$ of the frequency, and we can thus also assess the actual error in the estimation, which we define as $|\langle f_B\rangle-f_{B\text{,sim}}|$ and plot its median in Fig.~\ref{fig:FigS4x}(d) for the same collection of estimations as used in Fig.~\ref{fig:FigS4x}(c). 
	
	These findings show first of all that the non-tracking scheme [black in (a)] is relatively ineffective and, among the methods using a physics-informed prior distribution, the adaptive-time scheme performs best.
	Furthermore, we see that the uncertainties $\sigma$ plotted in panel (c), which we use in the main text as a measure for the estimation error, indeed correlate with the median of actual errors, shown in (d) (as also investigated above).
	We note that the overall error can be as low as \SI{2.5}{\mega\hertz}, with $N=5$, which would translate to total estimation time $T_{\text{est}} = 5\cdot\SI{20}{\micro\second} = \SI{100}{\micro\second}$, where $\SI{20}{\micro\second}$ is the typical duration of a single free-induction decay experiment. 
	
	\clearpage
	\section{Physics-informed adaptive Bayesian tracking: setup limitations}
	
	In this section we discuss the limitations of our setup.
	We show experimental results where the adaptive-time scheme deviates from the expected behavior, affecting the statistics of the performance of the scheme. We explain how we accounted for them when presenting the performance of the scheme in the main text, and we also discuss potential underlying causes and propose possible solutions for future work.
	
	\subsection{Bias towards lower frequencies}
	
	\begin{figure}[h]
		\centering
		\includegraphics[width=\textwidth]{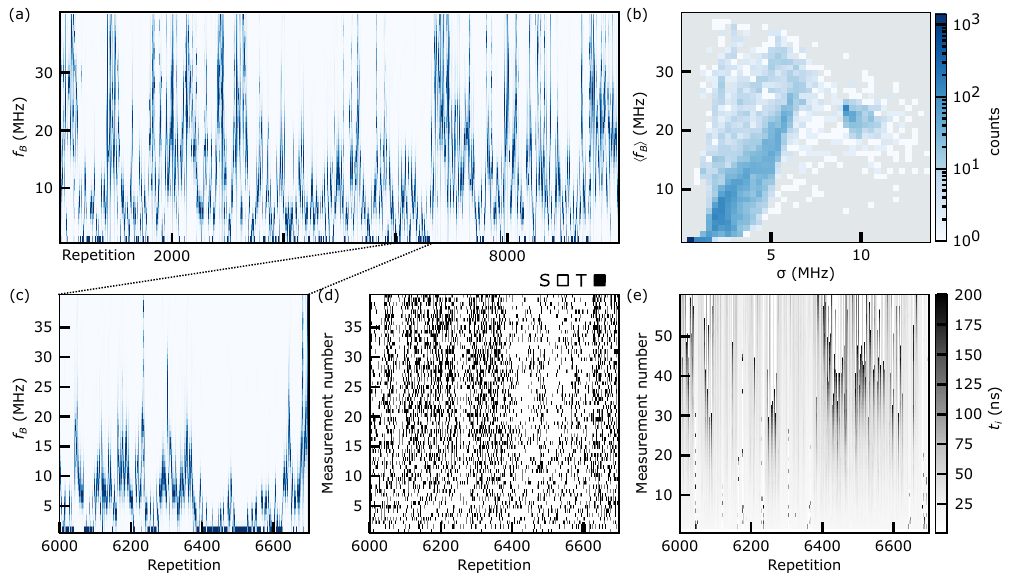}
		\caption{\textbf{Bias of the adaptive-time estimation scheme at low frequencies.}
			\textbf{(a)} An example of a complete adaptive-time estimation trace across 10,000 repetitions ($\approx \SI{30}{\second}$ of laboratory time), with $N=40$ and $T_{\rm op} = \SI{20}{\milli\second}$.
			\textbf{(b)} Histogram of the means and standard deviations $\{\langle f_B\rangle, \sigma_f\}$ of all repetitions in (a).
			\textbf{(c)} Zoom in on repetitions 6,000--6,700 from (a). 
			\textbf{(d)} Single-shot measurement outcomes used for the estimations during the repetitions shown in (c).
			\textbf{(e)} Sensing times used for all single-shot measurement across the different repetitions.
		}
		\label{fig:FigS5x}
	\end{figure}
	
	Fig.~\ref{fig:FigS5x}(a) shows the final distributions of a series of 10,000 adaptive-time estimations (approximately $\SI{30}{\second}$ of lab time) where we used $c =6$, $N = 40$, and $T_{\rm op} = \SI{20}{\milli\second}$.
	We immediately notice that there seems to be an unexpectedly large number of distributions that peak sharply in the first bin (the frequency range $0$--$\SI{1}{\mega\hertz}$).
	We corroborate this observation in Fig.~\ref{fig:FigS5x}(b), which shows a two-dimensional histogram of the final mean and standard deviation $\{\langle f_B\rangle, \sigma_f\}$ of all estimations shown in (a).
	We see that there is indeed a disproportionate number of estimated frequencies taking the lowest value of $\langle f_B \rangle \approx\SI{0.5}{\mega\hertz}$, all having an extremely narrow final distribution.
	We consider these estimations to be anomalies caused by the limitations of our setup, and thus filtered them out before performing the benchmarking presented in the main text.
	Since we thus reject a number of estimations with very small associated uncertainty, our comparison of the adaptive-time scheme to other methods in Fig.~4(c) of the main text can be seen as a worst-case benchmark for the experimentally implemented adaptive-time scheme.
	
	To investigate the cause of this behavior, we zoom in on a range of repetitions with a significant number of anomalous estimations, shown in Fig.~\ref{fig:FigS5x}(c), and we plot the corresponding single-shot measurement outcomes $m_i$ [Fig.~\ref{fig:FigS5x}(d)] and sensing times $t_i$ [Fig.~\ref{fig:FigS5x}(e)] used during each estimation.
	This allows us to make two observations:
	(i) The ranges where the estimated frequency is small correspond to white ``stripes" of singlet-biased data in Fig.~\ref{fig:FigS5x}(d), see, e.g., the repetitions in the interval [6400,6600].
	(ii) Inside these ranges, the sensing times $t_i$, shown in Fig.~\ref{fig:FigS5x}(e), increase much faster than elsewhere; eventually it reaches the maximum sensing time of $\SI{200}{\nano\second}$ allowed by the quantum controller, after which the time is reset to the user-defined value of $\SI{1}{\nano\second}$. 
	Below we will discuss several mechanisms we identified that could play a role in this behavior, and we give an outlook on possible ways to mitigate these issues.
	
	\textbf{Qubit dephasing.} 
	We recall that the sensing time is determined by the posterior distribution variance at each step as $t_i = 1/c\sigma_{i-1}$, where $c \approx 13$ was used in the experiments.
	This in fact helps the estimation being attracted to small $\langle f_B\rangle$ and $\sigma_f$ during the estimation cycle in the following way:
	Small posterior uncertainties $\sigma_i$ lead to longer separation times, which at some point become comparable with the qubit decoherence time $T_2$, on the order of $\approx \SI{100}{\nano\second}$.
	When $t_i \gtrsim T_2$ there is no information left in the measurement outcomes and the probabilities for measuring $m_i = \pm 1$ become independent of $t_i$.
	Since the scheme we use does not include a finite dephasing in the likelihood function, the absence of oscillations at larger times is in fact processed as correct information and can be treated as evidence for vanishing $f_B$ (depending on the saturation value of the singlet probability), yielding (i) an estimate with both very small $\langle f_B \rangle$ and $\sigma_f$ and (ii) a relatively quick divergence of $t_i$ during the estimation, as indeed seen in Fig.~\ref{fig:FigS5x}(e).
	
	A straightforward improvement in the estimation scheme would be to modify Eq.~(3) of the main text by adding a phenomenological dephasing time $T_2$ as follows
	\begin{equation}
		\begin{aligned}
			P_{\text{final}}\left(f_B\right) \propto P_0\left( f_B\right)  \prod_{i=1}^N\left[1+m_i\left(\alpha+\beta\, \textrm{e}^{-t_i/T_2} \cos \left(2 \pi f_B t_i\right)\right)\right].
		\end{aligned}
	\end{equation}
	The exponential factor will decrease the weight of the information gained at longer separation times, thus taking into account the decoherence of the qubit.
	
	\textbf{Residual exchange coupling.}
	The relatively high number of data points in the lowest frequency bin could also be partly attributed to a non-vanishing field along the $z$-axis of the qubit Bloch sphere at the sensing point deep in the $(1,1)$ region, due to residual exchange coupling.
	When the magnitude of the Overhauser field gradient becomes comparable to or smaller than the instantaneous exchange splitting, i.e., $h f_B \lesssim J(\varepsilon)$, then the free-induction decay precession on the Bloch sphere will no longer let the qubit evolve approximately along a meridian from the north pole (singlet) to the south pole (triplet) and back, but rather only partly reach the triplet state. This will thus bias the measurement outcomes towards more singlets in this low-Overhauser-field limit. Since the likelihood function we use for the Bayesian update assumes the rotations to be along the $x$-axis and thus does not account for residual exchange coupling, the bias towards more measured singlets results in a bias towards believing that the qubit does not rotate at all, thus contributing to confidence that the frequency is zero. As the Overhauser field becomes larger this biasing effect becomes less and less pronounced.
	
	The residual exchange may be reduced by reducing the tunnel coupling between the two dots during FID. Alternatively, the estimation protocol could be modified by considering the residual exchange when the qubit frequency is estimated. 
	
	\textbf{Gaussian approximation.}
	The way we convert each final posterior distribution to a Gaussian, as input for the physics-informed evolution of the distribution, can contribute to artificial narrowing of the distribution for small $\langle f_B\rangle$.
	Indeed, an underlying probability distribution for $f_B$ that has significant weight at both positive and negative frequencies will yield a distribution for $|f_B|$ on the quantum controller that is in fact narrowed, up to a factor $2$.
	The $\sigma_f$ extracted as input for the FP evolution can thus be smaller than the actual uncertainty in the underlying distribution, and this artificial reduction will contribute to the bias toward small $\{\langle f_B\rangle, \sigma_f \}$ as described above.
	
	For future experiments, one could try to derive an improved version of the FP equation, that takes the indiscernibility of the sign of $f_B$ into account, not producing artificial narrowing of $\sigma$ for small frequencies.
	However, one can argue that for any practical purpose (e.g., using the Overhauser gradient as a coherent control axis) the regime of very small $f_B$ should be avoided anyway, and the pragmatic way to mitigate this issue is thus simply to discard estimations that yield an $\langle f_B \rangle$ below some user-defined threshold.
	We note that this ``sign problem'' also plays a role in the numerical simulations we performed, in contrast to the qubit dephasing and residual exchange coupling discussed above, which we set to be absent in the simulations.
	
	\subsection{Numerical errors on the quantum controller hardware}
	
	\begin{figure}[ht]
		\centering
		\includegraphics[width=\textwidth]{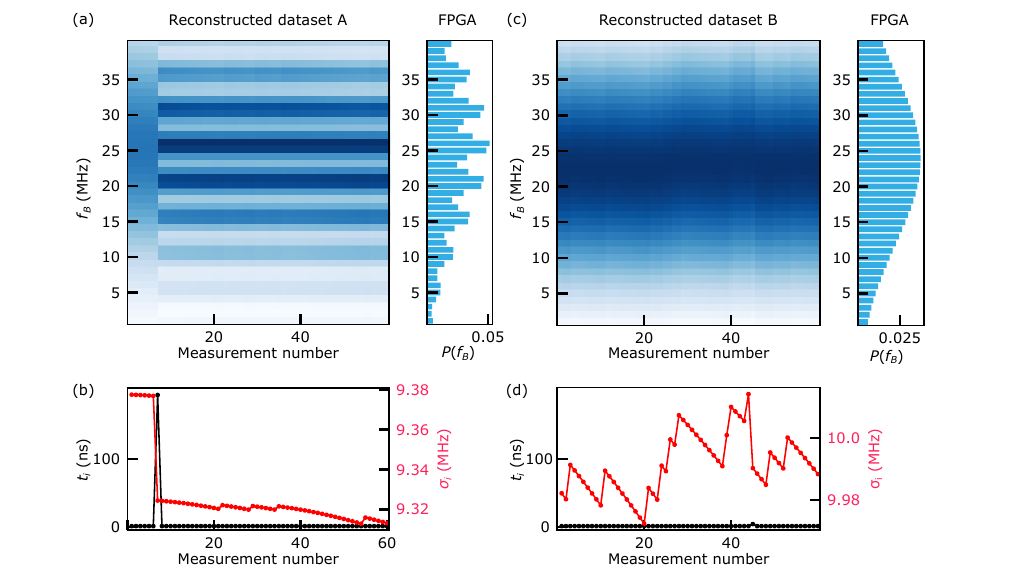}
		\caption{\textbf{Examples of potential numerical errors.}
			\textbf{(a)} (left) Reconstructed (by post processing) Bayesian update of a distribution that becomes multi modal.
			(right) The corresponding experimental posterior as computed on the quantum controller.
			\textbf{(b)} Sensing time and standard deviation for every single-shot outcome in the estimation run.
			\textbf{(c)} (left) Reconstructed Bayesian update of a distribution that remains almost unchanged during estimation.
			(right) The corresponding final posterior as computed on the quantum controller. \textbf{(d)} sensing time and standard deviation for every single-shot outcome in the estimation run.
		}
		\label{fig:FigS6x}
	\end{figure}
	
	Finally, we note that in some cases $\sigma_{i}$ becomes sufficiently small or large to cause errors on the FPGA-powered quantum controller, because of its available numerical accuracy, leading to presumably incorrect estimations.
	In Figure~\ref{fig:FigS6x} we show two examples of unexpected behavior during an estimation that we attribute to an error on the quantum controller.
	In (a,c) we show in the right panel the final distributions of the two estimations as stored on the quantum controller and in the left panel we present the reconstructed evolution of the distribution during the estimation, as explained above.
	In (b,d) we show the corresponding series of $t_i$ and $\sigma_i$ during the estimation, resulting from the reconstruction.
	In dataset A [Fig~\ref{fig:FigS6x}(a,b)] we see that the sensing time $t_i$ suddenly jumps to a large value ($\approx\SI{200}{\nano\second}$) and then drops back again to a very small value ($\approx\SI{1}{\nano\second}$).
	This results in a posterior that has many peaks, due to the rapid oscillation of the likelihood function $P\left(m_i | f_B \right)$ when $t_i$ is large. When waiting $\SI{20}{\milli\second}$ between estimations, such upward jumps in $t_i$ over $\SI{100}{\nano\second}$ happen for $30\%$ of estimations.
	In dataset B [Fig~\ref{fig:FigS6x}(c,d)] the sensing time [Fig~\ref{fig:FigS6x}(d)] almost does not change at all from $\SI{1}{\nano\second}$ so that the final posterior distribution is almost equal to the initial prior distribution. This seems to happen much less frequently, with only about $1\%$ of estimations (with $\SI{20}{\milli\second}$ waiting time between estimations) having values of $t_i$ that are all smaller than $\SI{5}{\nano\second}$.
	We note that in both datasets the final distributions end up having relatively large variances. When the waiting time in between estimations is $\SI{1}{\milli\second}$ or $\SI{5}{\milli\second}$, the only instances where $t_i$ suddenly jumps up is when the standard deviation is very low, and there are no estimations where $t_i$ are all smaller than $\SI{5}{\nano\second}$.
	
	It is only in the $T_{\text{op}}=\SI{20}{\milli\second}$ case where the starting standard deviation at each estimation is above $\SI{8.9}{\mega\hertz}$, indicating that numerical overflows are the likely cause.
	Indeed, overflow errors in the quantum controller are expected to happen for variances outside the range of standard deviations $1/(8\sqrt{10})\, \SI{}{\mega\hertz} \approx \SI{40}{\kilo\hertz}< \sigma < \sqrt{80}\,\SI{}{\mega\hertz}\approx \SI{8.9}{\mega\hertz}$. However, the adaptive separation times produced during overflow errors are not retrievable through post-processing, preventing a definitive attribution of these errors to the aforementioned overflow issues. Nevertheless, for the two examples depicted in the figures above, we observe that the reconstructed standard deviations (which are in good agreement with the final posterior standard deviations found by the quantum controller) are greater than the $\SI{8.94}{\mega\hertz}$ threshold capable of causing the quantum controller to overflow.
	
	A possible solution for future works when evaluating the standard deviation on the FPGA is to count the number of zeros in the mantissa of the fixed point number [limited to the range $[-8,8)$] and choose a different conversion factor accordingly, at the expense of added program complexity.
	
	\nocite{*}
	
	\bibliography{suppl_my_bibliography}